\documentclass[draft, preprintnumbers, nofootinbib, preprint, 12pt]{revtex4}
\usepackage{amsmath,amssymb,bm}
\begin{document}
{~}
\vspace{3cm}
\title{Multi-Black Rings on Eguchi-Hanson Space
\vspace{1cm}
}
\author{
	Shinya Tomizawa\footnote{E-mail:tomizawa@sci.osaka-cu.ac.jp}
}
\affiliation{ 
Department of Mathematics and Physics,
Graduate School of Science, Osaka City University,
3-3-138 Sugimoto, Sumiyoshi, Osaka 558-8585, Japan
\vspace{3cm}
}

\preprint{
OCU-PHYS 287\ AP-GR 53}

\begin{abstract}
We construct new supersymmetric multi-black ring solutions on the Eguchi-Hanson base space as solutions of the five-dimensional minimal supergravity. The space-time has an asymptotically locally Euclidean time slice, i.e., it has the spatial infinity with the topology of the lens space $L(2;1)=\rm S^3/{\mathbb Z}_2$. The configurations of black rings are restricted by the requirement of the absence of a Dirac-Misner string everywhere outside horizons. Especially, in the case of two black rings, the solutions have the limit to a pair of rotating black holes with the horizon topology of $\rm S^3$.  
\end{abstract}

\date{\today}
\maketitle


\section{Introduction}
One of striking features of asymptotically flat black holes in five dimensions is that they admit event horizons with non-spherical topology~\cite{Cai,Helfgott,Galloway} in contrast to four dimensions~\cite{Hawking,hawking_ellis}. The black ring solutions with horizon topology $\rm S^1\times S^2$, which rotate in the $\rm S^1$ direction, were found by Emparan and Reall as solutions to the five-dimensional vacuum Einstein equations~\cite{Emparan}. This is the first example of asymptotically flat black hole solutions with such non-spherical horizon topology. Remarkably, within some range of the parameters, there are a black hole~\cite{Myers} and two black rings for the same values of the mass and the angular momentum, which means the violation of the black hole uniqueness~\cite{uni,Israel,Israel2,Bunting,Bunting2,Carter,Robinson,Robinson2,Mazur} known in four dimension (See Ref.\cite{Morisawa,MTY} about the discussion on the uniqueness of a black hole and a black ring solution.). Subsequently, by using solitonic techniques, another black ring solutions were found. The black ring solutions with a rotating two sphere were found by Mishima and Iguchi~\cite{MI} by the B\"acklund transformation (they were independently found by Figueras~\cite{F}), and moreover, ones with two angular momenta were constructed by Pomeransky and Sen'kov~\cite{Pomeransky} by using the inverse scattering method~\cite{solitonbook,Belinskii,Belinskii2,exactbook,Tomizawa,Tomizawa2,Tomizawa3,Tomizawa4,Iguchiboost,Pomeransky:2005sj,Koikawa,Azuma,EK}. Elvang and Figureas generated a black Saturn solution, which describes a spherical black hole surrounded by a black ring~\cite{EF}. Furthermore, Iguchi and Mishima also generated a black di-ring solution~\cite{MI3}. An orthogonal black di-ring solution was also constructed~\cite{Izumi,Elvangorthogonal}.

Based on the classification of the solutions in the five-dimensional minimal supergravity~\cite{Gauntlett3}, in addition to a black hole solution~\cite{BMPV}, several supersymmetric black ring solutions have been constructed. The point is that they have been constructed on any four dimensional hyper-K\"ahler base spaces, especially, the Gibbons-Hawking base space. Elvang {\it et.al.} found the first supersymmetric black ring solutions with asymptotic flatness on a four dimensional Euclid space~\cite{Elvang}. They also presented more general supersymmetric black ring solutions with three charges~\cite{Elvang2}. Gauntlett and Gutowski constructed concentric multi-black ring solutions on the base space~\cite{Gauntlett} and also generalized these solutions to black ring solutions with multiple charges~\cite{Gauntlett2}. The supersymmetric black ring solutions with a compactified extra dimension on the Taub-NUT base space were constructed~\cite{Bena,Bena2,Elvang3,Gaiotto}. See Refs.~\cite{BRreview} about the detail review of black ring solutions.

So far, most of people have much attention to asymptotically flat black hole solutions.
In four dimension, asymptotic flatness, which in general means the spacetime approaches a Minkowski spacetime at the infinity, is expected to be a good idealization of an isolated system. In higher dimensions, such an asymptotic Minkowski spacetime is considered to be realized if black holes are small enough compared with tension of the brane or the curvature radius of the bulk or the size of extra dimensions. 
However, in higher dimensional spacetimes, the asymptotic flatness admits a variety of rich structures in the sense that the curvatures vanish at the infinity.
In fact, higher dimensional black holes admit a variety of asymptotic structures, although the spacetimes become flat at the infinity. Kaluza-Klein black hole solutions~\cite{DM,GW,Rasheed,IM,IKMT,NIMT,TIMN,TIMN2,MINT} have the spatial infinity with a compactified extra dimension, i.e., the spacetime approaches a twisted $\rm S^1$ bundle over a four-dimensional Minkowski spacetime. Black hole solutions on the Eguchi-Hanson space~\cite{IKMT2} have the spatial infinity of topologically various lens spaces $L(2n;1)={\rm S}^3/{\mathbb Z}_{2n}$ ($n$:natural number). Since the latter black hole spacetimes have asymptotically locally Euclidean timeslice, the asymptotic structures of the spacetimes are locally isometric to a five-dimensional Minkowski spacetime. The supersymmetric black ring solutions with non-trivial asymptotic structure were constructed in Ref.~\cite{TIKM}. However, the properties of these black object solutions are considerably different from that of the black hole in asymptotic Minkowski spacetimes. For instance, the Kaluza-Klein black holes~\cite{IM,IKMT} 
and the black holes on the Eguchi-Hanson space~\cite{IKMT2} can have the horizon of lens spaces in addition to ${\rm S}^3$. The location of the black rings on the Eguchi-Hanson is restricted unlike the black ring on the Euclid space~\cite{Elvang}. 

In the five-dimensional Einstein-Maxwell theory with a positive cosmological constant, black hole solutions 
on the Euclid base space~\cite{London}, the Taub-NUT base space~\cite{IIKMMT} and 
the Eguchi-Hanson base space~\cite{IMK} were also constructed. In general, these black hole solutions are dynamical.
In particular, two-black hole solution on the Eguchi-Hanson space describes 
a non-trivial coalescence of black holes. 
In Refes. \cite{IMK,YIKMT}, the authors compared the two-black hole solution 
on the Eguchi-Hanson space with the two-black holes solution 
on the Euclid space~\cite{London}, 
and discussed how the coalescence of five-dimensional black holes depends on 
the asymptotic structure of the spacetime. 
Two black holes with the topology of $\rm S^3$ coalesce into a single black hole with the topology of the lens space $L(2;1)=\rm S^3/{\mathbb Z}_2$ in the case of 
Eguchi-Hanson space, while two black holes with the topology of $\rm S^3$ coalesce into a single black hole with the topology of $\rm S^3$ in the Euclid case. 
When the action has the Chern-Simon term, black holes can rotate~\cite{KS,MIKT}.

There is no reason to restrict ourselves to the spacetime which asymptotes to the Minkowski spacetime at the infinity. For example, in the context of Kaluza-Klein theory, in the presence of a lot Kaluza-Klein monopoles, the spatial infinity has a topological structure of a lens space $L(n,1)=S^3/{\mathbb Z}_n$. Hence in a spacetime with such a non-trivial spatial infinity, it is an important issue to study the properties of black hole and black ring solutions. The end of this article is to generalize the supersymmetric black ring solutions on the Eguchi-Hanson base space in Ref.~\cite{TIKM} to multi-black ring solutions and to investigate the features of the multi-black solutions. The construction of the solutions is based on the discussion in Ref.~\cite{Gauntlett3,TIKM}. We study what the possible configuration of the black rings is. The existence of more than one nut, which means an isolated fixed point of the action of one parameter family of an isometry, make it difficult to construct black ring solutions since Misner strings appears on the nuts. A flat space and the Euclidean self-dual Taub-NUT space has a single nut, while the Eguchi-Hanson space has two nut. However, as discussed in Ref.\cite{TIKM}, if we impose  the reflection symmetry on the locations and the shapes of black rings, we can also construct multi-black ring solutions on the Eguchi-Hanson space.

The remainder of this article is organized as follows. In Sec.\ref{sec:pre}, we give a brief review in Ref.\cite{Gauntlett3}. In Sec.\ref{sec:solution}, we construct new multi-black ring solutions on the Eguchi-Hanson space, which is the results based on the previous work~\cite{TIKM}. In Sec.\ref{sec:property}, we discuss the properties of these solutions, in particular, asymptotic structure, possible configurations of black rings, and the limit to black holes. We investigate the limit of the two-black ring solutions to two-black hole solutions. In Sec.\ref{sec:summary}, we summarize our results and give some discussions.


\section{Multi-black ring solutions}\label{sec:pre}
\subsection{Construction of solutions}
The bosonic sector of the five-dimensional minimal supergravity is the Einstein-Maxwell theory with a Chern-Simon term. Following the reference \cite{Gauntlett3},
all supersymmetric solution of five-dimensional minimal supergravity have a non-spacelike Killing vector field. In a region where the Killing vector field $\partial_t$ is timelike,
the metric and the gauge potential are given by
\begin{eqnarray}
ds^2=-H^{-2}(dt+\bm\omega)^2+Hds^2_{\cal B},\ {\bm A}=\frac{\sqrt{3}}{2}[H^{-1}(dt+\bm\omega)-\bm\beta],
\end{eqnarray}
respectively.  where the base space $ds^2_{\cal B}$ is a metric of an arbitrary hyper-K\"ahler space. In this article, we choose the Eguchi-Hanson space as the base space. The scalar function $H$, one-forms $\bm\omega$ and $\bm\beta$ on ${\cal B}$ are given by 
\begin{eqnarray}
\Delta H=\frac{4}{9}(G^+)^2,\quad dG^+=0,\quad d{\bm \beta}=\frac{2}{3}G^+.
\end{eqnarray}
Here, $\triangle$ is the Laplacian on ${\cal B}$ and the two-form $G^+$ is the self-dual part of the one-form $H^{-1}{\bm \omega}$, i.e., 
\begin{eqnarray}
G^+:=\frac{1}{2}H^{-1}(d\omega+*d\omega),
\end{eqnarray}
where $(G^+)^2:=\frac{1}{2}G_{mn}G^{mn}$ and $*$ is the Hodge dual operator on ${\cal B}$.

As is shown in Appendix A, the metric of the Eguchi-Hanson space in the Gibbons-Hawking coordinates is given by
\begin{eqnarray}
ds^2_{EH}=H_k(dr^2+r^2d\Omega_{S^2}^2)+H_k^{-1}\left(\frac{a}{8}d\psi+\bm\varphi\right)^2,\label{eq:EH}
\end{eqnarray}
where $d\Omega^2_{S^2}=d\theta^2+\sin^2\theta d\phi^2$. The harmonic function on the three-dimensional Euclid space $H_k$ takes the form of~\cite{IKMT}
\begin{eqnarray}
H_k&=&\frac{a}{8}\left(\frac{1}{\Delta_{a_1}}+\frac{1}{\Delta_{a_2}}\right).\label{eq:H_k}
\end{eqnarray}
The one-form $\bm\varphi$ is determined by the equation, ${\rm rot}\ {\bm \varphi}={\rm grad}\ H_k$, and it is explicitly written as
\begin{eqnarray}
\bm\varphi=\frac{a}{8}\left(\frac{(r\cos\theta+a_1)}{\Delta_{a_1}}+\frac{(r\cos\theta+a_2)}{\Delta_{a_2}}\right)d\phi.
\end{eqnarray}  
Here, the functions $\Delta_{a_i} (i=1,2)$ are defined as $\Delta_{a_i}=\sqrt{r^2+2a_ir\cos\theta+a_i^2}$ with $a_1=-a_2=-a$. The coordinates run the ranges $0 <r$, $0\le \theta <\pi$, $0\le \phi< 2\pi$ and $0\le \psi<4\pi$. $\partial_\psi$ is a Killing vector field with closed orbits on the base space and it has fixed points at the point sources of the harmonic function $H_k$. Such a fixed point of a Killing vector field is called a {\it nut}. It should be noted that if Eq.(\ref{eq:H_k}) is replaced by a harmonic function with a single nut, i.e., 
\begin{eqnarray}
H_k=\frac{a}{8}\frac{1}{\Delta_{a_1}},
\end{eqnarray}
the metric (\ref{eq:EH}) coincides with that of the four-dimensional Euclid space.

The functions $H$ and the one-form ${\bm \omega}$ can be solved explicitly if the Killing vector field $\partial_\psi$ is also a Killing vector field of the full five-dimensional spacetime, i.e.,  $H$ and ${\bm \omega}$ are independent of $\psi$ and furthermore, if the one-forms $\bm\omega$ and $\bm\beta$ can be written in the form 
\begin{eqnarray}
\bm\beta&=&\beta_0\left(\frac{a}{8}d\psi+\bm\varphi\right)+\tilde{\bm\beta}, \\
 \bm \omega&=&\omega_0\left(\frac{a}{8}d\psi+\bm\varphi\right)+\tilde{\bm\omega}.
\end{eqnarray}
Then, under these assumptions, the functions $H$, $\omega_0$ and $\beta_0$ are written as 
\begin{eqnarray}
       H&=&H_k^{-1}K^2+L,\\
       \omega_0&=&H_k^{-2}K^3+\frac{3}{2}H_k^{-1}KL+M,\\
       \beta_0&=&H_k^{-1}K,
\end{eqnarray}
where $K$, $L$ and $M$ are another harmonic functions on a three dimensional Euclid space. The one-forms $\tilde{\bm\omega}$ and $\tilde{\bm\beta}$ are determined by the equations
\begin{eqnarray}
d{\tilde{\bm \omega}}=* \left[ H_{k}dM-MdH_{k}+\frac{3}{2}(KdL-LdK)\right],\label{eq:tomega}
\end{eqnarray}
\begin{eqnarray}
d\tilde{\bm\beta}=-*dK,\label{eq:tbeta}
\end{eqnarray}
where $*$ denotes the Hodge dual on the three dimensional Euclid space.
Assume that all point sources of three harmonic functions $K,L$ and $M$ are also located on the $z$-axis of the three dimensional Euclid space, i.e.,
\begin{eqnarray}
K&=&k_0+\sum_i\frac{k_i}{\Delta_{R_i}},\label{eq:k}\\
L&=&l_0+\sum_i\frac{l_i}{\Delta_{R_i}},\label{eq:l}\\
M&=&m_0+\sum_i\frac{m_i}{\Delta_{R_i}},\label{eq:m}
\end{eqnarray}
with
\begin{eqnarray}
& &\Delta_{R_i}:=\sqrt{r^2+R_i^2+2R_ir\cos\theta},
\end{eqnarray}
where these harmonic functions have point sources at ${\bm r}={\bm R_i}:=(0,0,-R_i)$.
Then, substituting Eqs.(\ref{eq:k}),(\ref{eq:l}) and (\ref{eq:m}) into Eq.(\ref{eq:tomega}),(\ref{eq:tbeta}) and integrating them, we obtain the explicit forms of the one-forms $\tilde{\bm\omega}=\tilde\omega_\phi d\phi$ and $\tilde{\bm\beta}=\tilde \beta_\phi d\phi$ as follows
\begin{eqnarray}
\tilde\omega_\phi&=&\frac{a}{8}\sum_{i=1}^nm_i\frac{a_1(r\cos\theta+R_i)+r(r+R_i\cos\theta )}{(a_1-R_i)\Delta_{a_1}\Delta_{R_i}}+\frac{a}{8}\sum_{i=1}^nm_i\frac{a_2(r\cos\theta+R_i)+r(r+R_i\cos\theta )}{(a_2-R_i)\Delta_{a_2}\Delta_{R_i}}\nonumber\\
                 &+&\sum_{i,j\ge 1,i\not=j}^n\frac{3}{2}k_il_j\frac{R_i(r\cos\theta+R_j)+r(r+R_j\cos\theta )}{(R_i-R_j)\Delta_{R_i}\Delta_{R_j}}+\sum_{i=1}^n\left(\frac{3}{2}k_0l_i-\frac{3}{2}l_0k_i\right)\frac{r\cos\theta+R_i}{\Delta_{R_i}}\nonumber\\
                 & &-\frac{a}{8}\sum_{i=1}^2m_0\frac{r\cos\theta+a_i}{\Delta_{a_i}}+C_\omega,
\end{eqnarray}
\begin{eqnarray}
\tilde\beta_\phi=-\sum_ik_i\frac{r\cos\theta+R_i}{\Delta_{R_i}}+C_\beta,
\end{eqnarray}
where 
$C_\omega$ and $C_\beta$ are arbitrary constants. We put $k_0=0$ so that the metric component does not diverge at the spatial infinity. We can always put $C_\beta=0$ from the freedom of the gauge of the potential $\bm A$.
The $tt$-component of the metric behaves as $g_{tt}\simeq -l_0^{-2}$ for $r\to\infty$. To fix the normalization of the timelike Killing vector field at the infinity, we put $l_0=1$. As will be mentioned in the next subsection, $C_\omega$ and $m_i$ are determined by the requirement for the absence of Misner strings for given $k_i$ and $l_i$.


\subsection{Determination of $m_i$ and $C_\omega$}\label{sec:solution}

In this article, we construct multi-black ring solutions on the Eguchi-Hanson space such that there is no Dirac-Misner string everywhere in the space-time. The existence would yield closed timelike curves if one impose that there is no conical singularity. In our solutions the conditions $\tilde\omega_\phi(\theta=0)=0$ and $\tilde\omega_\phi(\theta=\pi)=0$ assure the absence of Misner strings.  Whether there exist multi-black ring solutions is essentially determined by the conditions for the absence of Dirac-Misner strings, namely, the existence of the parameters $C_\omega,m_0,\cdots, m_n$ satisfying the conditions $\tilde\omega_\phi(\theta=0)=\tilde\omega_\phi(\theta=\pi)=0$. For example, in the case of concentric black rings on a flat base space~\cite{Gauntlett}, the number of these independent conditions amounts to $n+1$ since the $z$-axis on the three-dimensional Euclid space in the Gibbons-Hawking coordinate are divided into $n+2$ intervals by a point source at ${\bm a_{1}}$ of the harmonics $H_k$ and $N$ point sources at ${\bm R_{1}},\cdots,{\bm R_{n}}$. In general, the conditions $ \tilde\omega_\phi(\theta=0)=\tilde\omega_\phi(\theta=\pi)=0$ for each interval give rise to $n+2$ independent equations to $C_\omega,m_0,\cdots,m_n$ satisfying all equations. Therefore, since it is assured that there exist these parameters, the locations of black rings are arbitrary. In contrast, in the case of the Eguchi-Hanson base space, the situation changes. Since there are two nuts located at point sources ${\bm a_1}$ and ${\bm a_2}$, the number of the intervals on the $z$-axis increase by one. Namely, the number of independent equations exceeds that of the parameters. Thus, in general, it is impossible to put black rings at arbitrary positions on the $z$-axis unlike concentric black ring solutions on the flat space. However, it should be noted that if we impose reflection symmetry on the location of the point sources ${\bm R_1},\cdots,{\bm R_n}$ on the $z$-axis, i.e., if we choose the parameters such that
\begin{eqnarray}
&&m_1=m_{2},\cdots,m_N=m_{2N},\ k_1=k_{2},\cdots,k_N=k_{2N},\ l_1=l_{2},\cdots,l_N=l_{2N}\nonumber\\
&& R_1=-R_{2},\cdots,R_N=-R_{2N}\label{eq:para1}
\end{eqnarray}
 if $n$ is even ($n=2N$ for a positive integer $N$),
and 
\begin{eqnarray}
&&m_1=m_{2},\cdots,m_{N-1}=m_{2N-2},\ k_1=k_{2},\cdots,k_{N-1}=k_{2N-2},\ l_1=l_{2},\cdots,l_{N-1}=l_{2N-2},\nonumber\\
&&R_1=-R_{2},\cdots,R_{N-1}=-R_{2N-2},\ R_{2N-1}=0 \label{eq:para2}
\end{eqnarray}
if $n$ is odd ($n=2N-1$), the black rings can be located on arbitrary places on the $z$-axis except ${\bm r}={\bm a_1},{\bm a_2}$ since the number of independent equations coincides with that of the parameters. Next we consider black ring solutions with the parameters satisfying (\ref{eq:para1}) or (\ref{eq:para2}).


\subsubsection{Black rings on an $ S^2$-bolt}
For simplicity,  we assume that all $(2N-1)$-black rings are located on an $\rm S^2$-bolt. Under the conditions (\ref{eq:para2}), we assume that $0<-R_1=R_2<\cdots<-R_{N-1}=R_{2N-2}<a,R_{2N-1}=0$. From Eqs. $\tilde\omega_\phi(\theta=0)=\tilde\omega_\phi(\theta=\pi)$=0, the parameters $C_\omega$, and $m_k\ (k=0,\cdots,2N-1)$ are obtained as  
\begin{eqnarray}
&&C_\omega=0,\label{eq:q1}\\
&&m_0=-\frac{6}{a}\left(k_{2N-1}+2\sum_{k=1}^{N-1}k_{2k-1}\right),\label{eq:q2}\\
&&m_{2i-1}=6\left(1-\frac{R_{2i-1}^2}{a^2}\right)\Biggl[k_{2i-1}-2\sum_{k=1}^{i-1}\frac{R_{2i-1}{\cal D}_{i,k}}{R_{2k-1}^2-R_{2i-1}^2}+2\sum_{k=i+1}^{N-1}\frac{R_{2k-1}{\cal D}_{k,i}}{R_{2i-1}^2-R_{2k-1}^2}
-\frac{{\cal D}_{N,i}}{R_{2i-1}}\Biggr],\label{eq:q3}\\
&&m_{2N-1}=6\left(k_{2N-1}+2\sum_{k=1}^{N-1}\frac{{\cal D}_{N,k}}{R_{2k-1}}\right),\label{eq:q4}
\end{eqnarray}
where ${\cal D}_{p,q}:=k_{2p-1}l_{2q-1}-k_{2q-1}l_{2p-1}$ and $i=1,\cdots,N-1$. On the other hand, if all $2N$-black rings are on an $\rm S^2$-bolt, assuming that $0<-R_1=R_2<\cdots<-R_N=R_{2N}<a$ under the conditions (\ref{eq:para1}), we obtain the parameters $C_\omega$ and $m_k\ (k=0,\cdots,2N)$ as
\begin{eqnarray}
&&C_\omega=0,\label{eq:qq1}\\
&&m_0=-\frac{12}{a}\left(\sum_{k=1}^{N}k_{2k-1}\right),\label{eq:qq2}\\
&&m_{2i-1}
=6\left(1-\frac{R_{2i-1}^2}{a^2}\right)\Biggl[k_{2i-1}-\sum_{k=1}^{i-1}\frac{2R_{2i-1}{\cal D}_{i,k}}{R_{2k-1}^2-R_{2i-1}^2}+\sum_{k=i+1}^{N}\frac{2R_{2k-1}{\cal D}_{k,i}}{R_{2i-1}^2-R_{2k-1}^2}\Biggr],\label{eq:2Nm}\label{eq:qq3}\\
&&m_{2N-1}
=6\left(1-\frac{R_{2N-1}^2}{a^2}\right)\left[k_{2N-1}-\sum_{k=1}^{N-1}\frac{2R_{2N-1}{\cal D}_{N,k}}{R_{2k-1}^2-R_{2N-1}^2}\right].\label{eq:qq4}
\end{eqnarray}

\subsubsection{Black rings outside an $S^2$-bolt}
Next we assume that all black rings are located outsides an $\rm S^2$-bolt. In the case of $(2N-1)$-black rings, under the assumption  $a<-R_1=R_2<\cdots<-R_{N-1}=R_{2N-2},R_{2N-1}=0$ and the conditions (\ref{eq:para2}), Eqs. $\tilde\omega_\phi(\theta=0)=\tilde\omega_\phi(\theta=\pi)=0$ determine the parameters $C_\omega$ and $m_k\ (k=0,\cdots,2N-1)$ as follows 
\begin{eqnarray}
&&C_\omega=0,\label{eq:p1}\\
&&m_0=-\frac{6}{a}\left(2\sum_{k=1}^{N-1}k_{2k-1}+k_{2N-1}\right),\label{eq:p2}\\
&&m_{2i-1}=-6\left(1-\frac{R_{2i-1}^2}{a^2}\right)\nonumber\\
&&{\hspace{1cm}}\times \biggl[\frac{ak_{2i-1}}{R_{2i-1}}+\sum_{k=1}^{i-1}\frac{2a{\cal D}_{i,k}}{R_{2i-1}^2-R_{2k-1}^2}-\frac{2a}{R_{2i-1}}\sum_{k=i+1}^{N-1}\frac{R_{2k-1}{\cal D}_{i,k}}{R_{2i-1}^2-R_{2k-1}^2}-\frac{a{\cal D}_{N,i}}{R_{2i-1}^2}\biggr],\label{eq:p3}\\
&&m_{2N-1}=6\left(2\sum_{k=1}^{N-1}\frac{{\cal D}_{N,k}}{R_{2k-1}}+k_{2N-1}\right).\label{eq:p4}
\end{eqnarray}
In the case of $2N$-black rings, assuming that $a<-R_1=R_2<\cdots<-R_N=R_{2N}$ under (\ref{eq:para1}), we obtain the parameters $C_\omega$ and  $m_{k}\ (k=0,\cdots,2N)$ as
\begin{eqnarray}
&&C_\omega=0,\label{eq:pp1}\\
&&m_0=-\frac{12}{a}\left(\sum_{k=1}^Nk_{2k-1}\right),\label{eq:pp2}\\
&&m_{2i-1}=-6\left(1-\frac{R_{2i-1}^2}{a^2}\right)\nonumber\\
&&  \hspace{1cm}\times \biggl[\sum_{k=1}^{i-1}\frac{2a{\cal D}_{i,k}}{R_{2i-1}^2-R_{2k-1}^2}+\frac{ak_{2i-1}}{R_{2i-1}}-\frac{2a}{R_{2i-1}}\sum_{k=i+1}^N\frac{R_{2k-1}{\cal D}_{i,k}}{R_{2i-1}^2-R_{2k-1}^2}\biggr],\label{eq:pp3}\\
&&m_{2N-1}=-6\left(1-\frac{R_{2N-1}^2}{a^2}\right)\left(\sum_{k=1}^{N-1}\frac{2a{\cal D}_{N,k}}{R_{2N-1}^2-R_{2k-1}^2}+\frac{ak_{2N-1}}{R_{2N-1}}\right),\label{eq:pp4}
\end{eqnarray}
where $i=1,\cdots,N-1$.

\subsubsection{Black rings on and outside an $S^2$-bolt}
Finally, we consider the case where there exists black rings on and outside an $\rm S^2$-bolt in the presence of more than two black rings.  For instance, in the case of odd black rings, i.e., $0<-R_1=R_2<\cdots<-R_l=R_{2l}<a<-R_{l+1}=R_{2(l+1)}<\cdots<-R_{N-1}=R_{2N-2},R_{2N-1}=0$, the parameters $C_\omega,m_{k}=m_{2k}\ (k=1,\cdots,l)$ are given by Eqs.(\ref{eq:q1})-(\ref{eq:q3}), and $m_{\hat k}=m_{2\hat k}\ (\hat k=l+1,\cdots,N-1)$ and $m_{2N-1}$ are given by Eqs.(\ref{eq:p3})-(\ref{eq:p4}). 

In the case of even black rings, i.e., $0<-R_1=R_2<\cdots<-R_l=R_{2l}<a<-R_{l+1}=R_{2(l+1)}<\cdots<-R_{N}=R_{2N}$, the parameters $C_\omega,m_{k}=m_{2k}\ (k=1,\cdots,l)$ and $m_{\hat k}=m_{2\hat k}\ (\hat k=l+1,\cdots,N)$ are given by Eqs.(\ref{eq:qq1})-(\ref{eq:qq3}) and Eqs.(\ref{eq:pp3})-(\ref{eq:pp4}), respectively.

\subsection{A single black ring}
In the special case of $n=1$, this solution coincides with a single black ring solution constructed in the previous article~\cite{TIKM}. Here we give the short review about the previous work. The parameters $C_\omega$ $m_{0}$ and $m_{1}$ are given by
\begin{eqnarray}
C_\omega=0,\quad m_0=-\frac{6k_1}{a},\quad m_1=6k_1. 
\end{eqnarray}
Then the metric takes the following form
\begin{eqnarray}
ds^2&=&-H^{-2}\left[d t+\omega_0\left(\frac{a}{8}d\psi+\varphi_\phi d\phi\right)+\tilde\omega_\phi d\phi\right]^2\nonumber\\
 &&+H\left[H_k(dr^2+r^2d\Omega_{\rm S^2}^2)+H_k^{-1}\left(\frac{a}{8}d\psi+\varphi_\phi d\phi\right)^2\right],
\end{eqnarray}
where $d\Omega_{\rm S^2}^2=d\theta^2+\sin^2\theta d\phi^2$. The coordinates $r,\psi,\phi,\theta$ run the ranges
\begin{eqnarray}
r>0,\ 0\le \psi\le 4\pi,\ 0\le \phi\le 2\pi,\ 0\le\theta\le\pi.
\end{eqnarray}
The five functions $H_k,H,\omega_0,\tilde\omega_\phi$ and $\varphi_\phi$ are given by 
\begin{eqnarray}
H_k=\frac{a}{8}\left(\frac{1}{\Delta_a}+\frac{1}{\Delta_{-a}}\right),\label{eq:Hk}
\end{eqnarray}
\begin{eqnarray}
H=1+\frac{l_1}{r}+\frac{8k_1^2\Delta_{a}\Delta_{-a}}{ar^2(\Delta_{a}+\Delta_{-a})},
\end{eqnarray}
\begin{eqnarray}
\omega_0=2k_1\left(-\frac{3}{a}+\frac{3}{r}+\frac{6(l_1+r)\Delta_{a}\Delta_{-a}}{ar^2(\Delta_{a}+\Delta_{-a})}+\frac{32k_1^2\Delta_{a}^2\Delta_{-a}^2}{a^2r^3(\Delta_{a}+\Delta_{-a})^2}\right),
\end{eqnarray}
\begin{eqnarray}
\tilde\omega_\phi&=&\frac{3k_1}{4\Delta_{a}\Delta_{-a}}[(r+a)(\Delta_{-a}-\Delta_{a})+((r+a)(\Delta_{-a}+\Delta_{a})-2\Delta_{-a}\Delta_{a})\cos\theta],\label{eq:tilomega}
\end{eqnarray}
\begin{eqnarray}
\varphi_\phi=\frac{a}{8}\left(\frac{a(\Delta_{-a}-\Delta_{a})+r(\Delta_{a}+\Delta_{-a})\cos\theta}{\Delta_{a}\Delta_{-a}}\right).\label{eq:varphi}
\end{eqnarray}
It is noted that our solutions have three independent parameters $l_1,k_1$ and $a$, where $k_1$ and $l_1$ are related to the dipole charge $q$ of the black ring and the total electric charge $Q_e$ by $k_1=-q/2$ and $al_1=4G_5Q_e/(\sqrt{3}\pi)-q^2$, and $a$ is the radius of the $\rm S^2$-bolt on the Eguchi-Hanson space. To avoid the existence of CTCs (closed timelike curves) outside the event horizon, we impose the following conditions on these parameters
\begin{eqnarray}
\quad k_1<0,\quad l_1>-4k_1.
\end{eqnarray}
It should be noted that we can choose the origin of the three-dimensional Euclid space ${\mathbb E}^3$ in the Gibbons-Hawking coordinate such that that $\bm{a}_1=-\bm{a}_2=(0,0,a)$ and $R_1=0$ without loss of generality. For example, if we shift the origin so that $\bm{a}_1=0$, we need change $\Delta_{-a}$, $\Delta_{a}$ and $r$ in Eqs.(\ref{eq:Hk})-(\ref{eq:tilomega}) into $r$, $\Delta_{2a}$ and $\Delta_{a}$, respectively, and moreover we need replace $r\cos\theta$, $\Delta_{-a}$ and $\Delta_{a}$ in Eq.(\ref{eq:varphi}) with $r\cos\theta+a$, $r$ and $\Delta_{2a}$, respectively.     

As shown in the previous work, the solutions have  the same two angular momentum components and the asymptotic structure on timeslices is asymptotically locally Euclidean. The $\rm S^1$-direction of the black ring is along the equator on a $\rm S^2$-bolt on the Eguchi-Hanson space since the horizon is located at $r=0$. There are isometries acting on the $\rm S^2$-bolt on the Eguchi-Hanson space. Using these isometries, we can set these two nuts on other poles on the $\rm S^2$-bolt. Therefore we can put a black ring along the other equator. Namely, we can construct a black ring along arbitrary equators on the $\rm S^2$ bolt.

\subsection{Two black rings}
Now we consider the solutions with a pair of black rings located on an $\rm S^2$-bolt ($0<-R_1=R_2<a$). In this case, the parameters $C_\omega$, $m_{0}$ and $m_{1}$ are given by
\begin{eqnarray}
C_\omega=0,\quad m_0=-\frac{12k_1}{a},\quad m_1=m_2=6k_1\left(1-\frac{R_1^2}{a^2}\right). 
\end{eqnarray}
Then the metric functions take the following form
\begin{eqnarray}
H=1+l_1\left(\frac{1}{\Delta_{R_1}}+\frac{1}{\Delta_{-R_1}}\right)+k_1^2H_k^{-1}\left(\frac{1}{\Delta_{R_1}}+\frac{1}{\Delta_{-R_1}}\right)^2,
\end{eqnarray}
\begin{eqnarray}
\omega_0&=&-\frac{12k_1}{a}-\frac{6k_1 R_1^2}{a^2}\left(\frac{1}{\Delta_{R_1}}+\frac{1}{\Delta_{-R_1}}\right)+k_1^3H_k^{-2}\left(\frac{1}{\Delta_{R_1}}+\frac{1}{\Delta_{-R_1}}\right)^3\nonumber\\
&&+\frac{3k_1}{2}H_k^{-1}(1+4H_k)\left(\frac{1}{\Delta_{R_1}}+\frac{1}{\Delta_{-R_1}}\right)+\frac{3k_1l_1}{2}H_k^{-1}\left(\frac{1}{\Delta_{R_1}}+\frac{1}{\Delta_{-R_1}}\right)^2,
\end{eqnarray}
\begin{eqnarray}
\tilde\omega_\phi&=&\frac{3ak_1}{4}\left(1-\frac{a^2}{R_1^2}\right)\biggl(-\frac{r^2+r(R_1-a)\cos\theta-aR_1}{(a+R_1)\Delta_{-a}\Delta_{R_1}}+\frac{r^2-r(R_1+a)\cos\theta+aR_1}{(R_1-a)\Delta_{-a}\Delta_{-R_1}}\nonumber\\
                & &+\frac{r^2+r(R_1+a)\cos\theta+aR_1}{(a-R_1)\Delta_{a}\Delta_{R_1}}+\frac{r^2+r(-R_1+a)\cos\theta-aR_1}{(a+R_1)\Delta_{a}\Delta_{-R_1}}  \biggr)\nonumber\\
                &&+\frac{3(k_1l_2-k_2l_1)}{4R_1}\frac{r^2-R_1^2}{\Delta_{R_1}\Delta_{-R_1}}\nonumber\\
                &&+\frac{3k_1}{2}\left(-\frac{r\cos\theta +R_1}{\Delta_{R_1}}-\frac{r\cos\theta -R_1}{\Delta_{-R_1}}+\frac{r\cos\theta -a}{\Delta_{-a}}+\frac{r\cos\theta +a}{\Delta_{a}}\right).
\end{eqnarray}
As will be explained later, the necessary and sufficient conditions for the absence of CTCs outside the horizons of two black rings are 
\begin{eqnarray}
k_1<0,\quad l_1>-4k_1\sqrt{1-\frac{R_1^2}{a^2}}.
\end{eqnarray}

Next, we consider the solutions with a pair of black rings outside an $\rm S^2$-bolt ($a<-R_1=R_2$). The parameters $C_\omega$, $m_{0}$ and $m_{1}$ are given by
\begin{eqnarray}
m_0=-\frac{12k_1}{a},\quad m_1=m_2=-\frac{6k_1a}{R_1}\left(1-\frac{R_1^2}{a^2}\right).
\end{eqnarray}
Then the metric functions are written in the form
\begin{eqnarray}
H=1+l_1\left(\frac{1}{\Delta_{R_1}}+\frac{1}{\Delta_{-R_1}}\right)+k_1^2H_k^{-1}\left(\frac{1}{\Delta_{R_1}}+\frac{1}{\Delta_{-R_1}}\right)^2,
\end{eqnarray}
\begin{eqnarray}
\omega_0&=&-\frac{12k_1}{a}+6k_1\left(\frac{R_1}{a}-\frac{a}{R_1}\right)\left(\frac{1}{\Delta_{R_1}}+\frac{1}{\Delta_{-R_1}}\right)+k_1^3H_k^{-2}\left(\frac{1}{\Delta_{R_1}}+\frac{1}{\Delta_{-R_1}}\right)^3\nonumber\\
     & &+\frac{3k_1l_1}{2}H_k^{-1}\left(\frac{1}{\Delta_{R_1}}+\frac{1}{\Delta_{-R_1}}\right)^2+\frac{3k_1}{2}H_k^{-1}\left(\frac{1}{\Delta_{R_1}}+\frac{1}{\Delta_{-R_1}}\right),
\end{eqnarray}

\begin{eqnarray}
\tilde\omega_\phi&=&-\frac{3a^2k_1}{4R_1}\left(1-\frac{a^2}{R_1^2}\right)\biggl(-\frac{r^2+r(R_1-a)\cos\theta-aR_1}{(a+R_1)\Delta_{-a}\Delta_{R_1}}+\frac{r^2-r(R_1+a)\cos\theta+aR_1}{(R_1-a)\Delta_{-a}\Delta_{-R_1}}\nonumber\\
                & &+\frac{r^2+r(R_1+a)\cos\theta+aR_1}{(a-R_1)\Delta_{a}\Delta_{R_1}}+\frac{r^2+r(-R_1+a)\cos\theta-aR_1}{(a+R_1)\Delta_{a}\Delta_{-R_1}}  \biggr)\nonumber\\
                &&+\frac{3(k_1l_2-k_2l_1)}{4R_1}\frac{r^2-R_1^2}{\Delta_{R_1}\Delta_{-R_1}}\nonumber\\
                &&+\frac{3k_1}{2}\left(-\frac{r\cos\theta +R_1}{\Delta_{R_1}}-\frac{r\cos\theta -R_1}{\Delta_{-R_1}}+\frac{r\cos\theta -a}{\Delta_{-a}}+\frac{r\cos\theta +a}{\Delta_{a}}\right).
\end{eqnarray}
The conditions for the absence of CTCs outside the horizons of two black rings are 
\begin{eqnarray}
k_1<0,\quad l_1>\frac{4k_1a}{R_1}\sqrt{\frac{R_1^2}{a^2}-1}.
\end{eqnarray}

As will be shown later, in both two black ring solutions, the horizons are located at ${\bm r}=\pm{\bm R_1}$.
It is noted that in addition to three parameters $k_1$, $l_1$ and $a$, both of two black ring solutions have an additional parameter $R_1$. This fact means that unlike a single black ring solution, the black rings can be located at arbitrary places on the $z$-axis in the Gibbons-Hawking coordinate as far as two black rings are located so that they have reflection symmetry about the origin in the Gibbons-Hawking coordinates.

\section{Properties}\label{sec:property}

\subsection{Asymptotic structure}
To study the asymptotic structure of the solutions, we introduce a new coordinate $\tilde r^2:=ar$. The asymptotic form of the metric for $\tilde r\to\infty$ becomes
\begin{eqnarray}
ds^2&\simeq&-dt^2+d\tilde r^2+\frac{\tilde r^2}{4}\left[\left(\frac{d\psi}{2}+\cos\theta d\phi\right)^2+d\theta^2+\sin^2\theta d\phi^2\right],\label{eq:asy}
\end{eqnarray}
where it is noted that the $\tilde r$ constant surface can be regarded as a Hopf bundle, i.e., a twisted $\rm S^1$ bundle over a $\rm S^2$ base space. If $d\psi/2$ is replaced by $d\psi$, this coincides with the metric of the five-dimensional Minkowski spacetime. In other word, the twisted fiber in Eq.(\ref{eq:asy}) has  half of the periodicity of $\rm S^3$, which implies that the time slices is asymptotically locally Euclidean, i.e., the spatial infinity has the topological structure of the lens space $L(2;1)=\rm S^3/{\mathbb Z}_2$.
Here we use new angular variables $\tilde \phi:=(2\phi+\psi)/4$, $\tilde \psi:=(-2\phi+\psi)/4$ and $\Theta:=\theta/2$. Then the asymptotic form can be rewritten as
\begin{eqnarray}
ds^2\simeq -dt^2+d\tilde r^2+ \tilde r^2(d\Theta^2+\cos^2\Theta d\tilde\phi^2+\sin^2\Theta d\tilde\psi^2).
\end{eqnarray}
The total mass and the total angular momenta with respect to $\partial_{\tilde \phi}$ and $\partial_{\tilde\psi}$ are obtained as
\begin{eqnarray}
&&M_{ADM}=\frac{\sqrt{3}}{2}Q_e=\frac{3\pi}{8G}\left[4\left(\sum_{i}k_i\right)^2+a\left(\sum_il_i\right)\right],\\
&&J_{\tilde \phi}=J_{\tilde \psi}=-\frac{\pi}{4G}\left[4\left(\sum_{i}k_i\right)^3+\frac{3}{2}a\left(\sum_{i}k_i\right)\left(\sum_{i}l_i\right)+\frac{a^2}{4}\left(\sum_{i}m_i\right)\right].
\end{eqnarray}
The mass and the electric charge satisfy the BPS condition. Two angular momenta are equal in contrast to the concentric multi-black ring solutions on a flat space~\cite{Gauntlett,Gauntlett2}.


\subsection{Near-Horizon and regularity}
Here we investigate the near-horizon geometry of our solutions. From the reflection symmetry of black rings, we consider only the neighborhood of the point sources ${\bm r}={\bm R}_{2i}\ (i=1,2,\dots)$.
Let us shift a origin of the three-dimensional Euclid space ${\mathbb E}^3$ in the Gibbons-Hawking coordinate so that one of nuts, ${\bm r}={\bm a}_1$ is located on the origin, i.e., ${\bm a}_1=0$. We define new coordinates $(x,y,\hat\phi,\hat\psi)$ as
\begin{eqnarray}
r=-(a+R_{2i})\frac{x+y}{x-y},\quad \cos\theta=-1+2\frac{1-x^2}{y^2-x^2}, \quad \phi=\hat\phi-\hat\psi,\quad \psi=\hat\phi+\hat\psi.
\end{eqnarray}
As will be explained below, $y\to\infty$ corresponds to event horizons.
Moreover, let us introduce new coordinates $(z,\zeta)$ defined by
\begin{eqnarray}
y=-\frac{P}{z},\quad x=\cos\zeta,
\end{eqnarray}
where $P$ is some constant with dimension of length. Since the metric is apparently singular at $z=0$, we introduce new coordinates $dt=dv-\sum_{k=0}^2(B_k/z^k)dz$, $d\hat\phi=d\hat\phi'-\sum_{l=0}^1(C_l/z^l)$ and $d\hat\psi=d\hat\psi'-\sum_{l=0}^1(C_l/z^l)$, where the constants $B_k (k=0,1,2)$ and $C_l (l=0,1)$ are suitably chosen: the constants $B_2$, $C_1$ and $B_0$ are chosen to cure the divergences $1/z$ in $g_{\hat\psi' z}$, $1/z^2$ and $1/z$ in $g_{zz}$, respectively; the constants $C_0$ and $B_0$ are determined so that $g_{zz}=O(z)$ for $z\to 0$.
In the neighborhood of the $2i$-th point source on the $\rm S^2$-bolt, i.e., ${\bm r} \simeq {\bm R}_{2i}$ ($a>|{\bm R}_{2i}|$), the metric behaves as
\begin{eqnarray}
ds^2&\simeq&g_{vz}^{(0)}dvdz+2g_{z\hat\phi'}^{(0)}dzd\hat\phi'+2g_{z\hat\psi'}^{(0)}dzd\hat\psi'\nonumber\\
     & &+\frac{a^2(3l_{2i}^2-8k_{2i}m_{2i})}{64k_{2i}^2}d\phi_2^2+k_{2i}^2\left[d\zeta^2+\sin^2\zeta d\phi_1^2\right],
\end{eqnarray}
where the angular coordinates $\phi_1$ and $\phi_2$ are defined as $\phi_1=\hat\phi'-\hat\psi'=\hat\phi-\hat\psi=\phi$ and $\phi_2=\hat\phi'+\hat\psi'$. They run the ranges of $0\le \phi_1 \le2\pi$ and $0\le \phi_2 \le 4\pi$, respectively.
On the other hand, near the $2i$-th point source outside the $\rm S^2$-bolt, ${\bm r}\simeq{\bm R}_{2i}$ ($a<|{\bm R}_{2i}|$), the metric behaves as
\begin{eqnarray}
ds^2&\simeq&g_{vz}^{(0)}dvdz+2g_{z\phi}^{(0)}dzd\hat\phi'+2g_{z\psi}^{(0)}dzd\hat\psi'\nonumber\\
     & &+
\frac{a^2(3l_{2i}^2-8k_{2i}m_{2i})}{16k_{2i}^2}d\hat\psi^{\prime 2}+k_{2i}^2\left[d\zeta^2+\sin^2\zeta d\hat\phi^{\prime 2}\right],
\end{eqnarray}
where $\hat\phi'$ and $\hat\psi'$ run the ranges of $0\le\hat\phi'\le2\pi$ and  $0\le\hat\psi'\le2\pi$, respectively. 
Since the explicit form of $g^{(0)}_{vz}, g^{(0)}_{z\hat\phi'}$ and $g^{(0)}_{z\hat\psi'}$ are unimportant, we do not write it here. Since the Killing vector field $V=\partial_v$ is null at $z=0$ and furthermore $V_\mu dx^\mu=g_{vz}^{(0)}dz$, the hypersurface $z=0$ is a Killing horizon, whose spatial topology is $\rm S^1\times S^2$. It is noted that the $\rm S^2$ of a black rings on an $\rm S^2$-bolt and a black ring outside an $\rm S^2$-bolt have the radius of $l_{2i}:=a\sqrt{(3l_{2i}^2-8k_{2i}m_{2i})/(16k_{2i}^2)}$ and the $\rm S^1$ of them have the radius of $k_{2i}$. 
Since the metric is analytic at the horizons and the nuts, there is no curvature singularity on and outside the event horizons.

\subsection{Absence of CTCs}
From the near horizon geometry, it is necessary that for each $i$ the following inequality are satisfied
\begin{eqnarray}
3l_i^2-8k_im_i>0,\quad k_i<0,\label{eq:ineq}
\end{eqnarray}
which is the condition that there is no CTC in the neighborhood of the horizons of all black rings.
To ensure the absence of CTCs everywhere outside the horizons, we must demand that the spatial part of the metric is positive definite. 
This metric is positive-definite if and only if the following two-dimensional matrix is positive-definite
\begin{eqnarray}
{\cal M}=\left(
\begin{array}{cc}
A& -C\\
-C& B
\end{array}
\right),
\end{eqnarray}
where $A,B$ and $C$ are given by
\begin{eqnarray}
A=H^3H_k^{-1}-\omega_0^2,\quad B=H^3H_kr^2\sin^2\theta-\tilde\omega_\phi^2,\quad C=\omega_0\tilde \omega_\phi.
\end{eqnarray}
Therefore, noting that $AB-C^2=H^3H_k(AH_k^2r^2\sin^2\theta-\tilde\omega_\phi^2)$, we obtain the condition
\begin{eqnarray}
{\cal M}>0 &\Longleftrightarrow& A>0, \quad AB-C^2>0\\
       &\Longleftrightarrow& AH_k^2r^2\sin^2\theta-\tilde\omega_\phi^2>0.
\end{eqnarray}
As a result, to show that the necessary conditions (\ref{eq:ineq}) are also sufficient for the absence of CTCs everywhere outside the horizons, it is enough to prove that $D:=AH_k^2r^2\sin^2\theta-\tilde\omega_\phi^2 >0$ there. In the previous work, we showed this in the case of a single black ring. We can also confirm numerically that this is also achieved for the special cases, i.e.,  the cases of two black rings. It is difficult to show that in general, these inequalities are also sufficient conditions for CTCs to vanish outside all horizons.

\subsection{Black hole limit}
Now we consider the limit of our solutions to black hole solutions. For simplicity, we concentrate on two black rings located on or outside an $\rm S^2$-bolt. 
Taking the limit ${\bm R}_1\to {\bm a}_1 ({\bm R}_2\to {\bm a}_2)$, we can obtain a pair of black holes on Eguchi-Hanson space:
\begin{eqnarray}
ds^2=-H^{-2}\left[dt+\omega_0\left(\frac{a}{8}d\psi+\varphi_\phi d\phi\right)\right]^2+Hds^2_{EH},
\end{eqnarray}
where in the limit, the functions $H$, $\omega_0$ and $\varphi_\phi$ take the form of  
\begin{eqnarray}
H=1+\frac{8k_1^2+al_1}{a}\left(\frac{1}{\Delta_{-a}}+\frac{1}{\Delta_{a}}\right),
\end{eqnarray}
\begin{eqnarray}
\omega_0&=&4k_1\frac{16k_1^2+3al_1}{a^2}\left(\frac{1}{\Delta_{-a}}+\frac{1}{\Delta_{a}}\right),
\end{eqnarray}
\begin{eqnarray}
\varphi_\phi=\frac{a}{8}\left(\frac{r\cos\theta-a}{\Delta_{-a}}+\frac{r\cos\theta+a}{\Delta_{a}}\right).
\end{eqnarray}
Note that in this limit, the functions $H$ and $\omega_0$ becomes harmonic functions on a three-dimensional Euclid space. Also note that the absence of a Dirac-Misner string is assured since $\tilde\omega_\phi\to 0$ in this limit. From the refection symmetry about $z=0$ on the $z$-axis in ${\mathbb E}^3$, we analyze only one of two point sources ${\bm r}=(0,0,a)$. We set up new spherical coordinates $(r,\theta,\phi)$ centered on this point. Then the metric written in the coordinates is singular at ${\bm r}=0$ since $g_{rr}$ behaves as $O(1/r^2)$ in the neighborhood of this point. To eliminate this divergence, we introduce new coordinates $(v,\psi')$ defined by $dt=dv+f(r)dr$, $d\psi=d\psi'+g(r)dr$, where the functions $f(r)$ and $g(r)$ are given by
\begin{eqnarray}
&&f(r)=\frac{l_1\sqrt{6k_1^2+al_1}}{2\sqrt{2}r^2}+\frac{192k_1^4+12k_1^2l_1(4a+l_1)+al_1^2(3a+2l_1)}{4\sqrt{2}al_1\sqrt{6k_1^2+al_1}r},\\
&&g(r)=\frac{\sqrt{2}k_1(16k_1^2+3al_1)}{al_1\sqrt{6k_1^2+al_1}}-\frac{k_1(16k_1^2+3al_1)(192k_1^4+6k_1^2l_1(8a-l_1)+a(3a-l_1)l_1^2)}{\sqrt{2}a^2l_1^3\sqrt{6k_1^2+al_1}^3}.
\end{eqnarray}
Then, near the point source of ${\bm r}=0$, the metric behaves as
\begin{eqnarray}
ds^2&\simeq& -\frac{2(8k_1^2+al_1)}{2\sqrt{2}l_1\sqrt{6k_1^2+al_1}}drdv+\left(k_1^2+\frac{al_1}{8}\right)(d\theta^2+\sin^2\theta d\phi^2)\nonumber\\
 & &+\left[k_1^2+\frac{al_1}{8}-\left(\frac{k_1(16k_1^2+3al_1)}{2(8k_1^2+al_1)}\right)^2\right](d\psi+(1+\cos\theta)d\phi)^2+{\cal O}(r).
\end{eqnarray}
In this coordinate system $(v,r,\theta,\phi,\psi')$, all of the metric components take finite values and are analytic. Hence the coordinate system is well-defined around this point. Since the Killing vector field  $V:=\partial_v$ becomes null at $r=0$ and it is hypersurface orthogonal, i.e., $V_\mu \propto dr$, the hypersurface $r=0$ is a Killing horizon.  
From the metric on the point source,
we find that the spatial topology of the horizon is a squashed $\rm S^3$. Therefore, in this limit, the solutions describe a pair of rotating black holes located at the north pole and the south pole on the $\rm S^2$-bolt. The angular velocities of the horizon vanish, although the total angular momenta evaluated at the spatial infinity do not vanish. Especially, if we restrict ourselves to the vanishing angular momentum case, the solutions exactly coincide with the black hole solutions in the case of equal masses obtained in Ref.~\cite{IKMT}.


\section{Summary and Discussions}\label{sec:summary}
In this article, we have constructed supersymmetric multi-black ring solutions on the Eguchi-Hanson base space as solutions of the five-dimensional minimal supergravity. The basic idea for constructing solutions is based on the programs to classify supersymmetric solutions of five-dimensional $N=1$ supergravity in Ref.~\cite{Gauntlett3}.  We have also investigated the properties of our black ring solutions for the configuration of more than one black rings. The space-time has the asymptotically locally Euclidean time slices, i.e., it has the spatial infinity with its topology the lens space $L(2,1)=\rm S^3/{\mathbb Z}_2$. If we assume all the point sources of the harmonics $K,L$ and $M$ are put on the $z$-axis in the Gibbons-Hawking space, the black ring solutions can be constructed. However, the configuration of the black rings are restricted so that they have the reflection symmetry about $z=0$. This results from the requirement of the absence of a Dirac-Misner string everywhere outside horizons. We have found that in the case of two black rings, the solutions have the limit to a pair of black holes with a squashed $\rm S^3$. Furthermore, in the vanishing limit of angular momenta, the solutions coincide with the black hole solutions with equal masses which was constructed in the previous work~\cite{IKMT}.  

In this article, assuming that the Killing vector field $\partial_\psi$ of the Eguchi-Hanson space is also the Killing vector field of the spacetime, we have constructed black ring solutions. Hence, there will exist general solutions which admit other configurations of black rings. In such solutions, the metric depends on the coordinate $\phi$. In particular, there are isometries acting on the $\rm S^2$-bolt on the Eguchi-Hanson space. Using these isometries, we can set these two nuts on other poles on the $\rm S^2$-bolt. Thus we can construct a black ring solution whose metric depends on $\phi$.

The Eguchi-Hanson space has two nuts but we can replace the harmonic function (\ref{eq:H_k}) by the more general form
\begin{eqnarray}
H_k=\sum_i\frac{Q_i}{\Delta_{a_i}},
\end{eqnarray} 
where $Q_i\  (i=1,2,\cdots)$ are constants. If all $Q_i$ take the same value $a/8$, the space is regular. If not so, conical singularities appear at nuts. There seems to be no solution to $\tilde\omega_\phi(\theta=0)=\tilde\omega_\phi(\theta=\pi)=0$ even if we impose some symmetry on the configurations and the arrangements of black rings. Hence there is no black ring solution in such a spacetime. This fact might suggest that in a spacetime with a lot of nuts, black rings cannot be produced.

It is interesting to see how the areas of black rings depend on the values of $R_1$ for the same masses and the same angular momenta.
 For simplicity, we consider the configuration of two black rings on a $\rm S^2$-bolt with its radius $a$ fixed. From Eq.(\ref{eq:2Nm}), the parameters $m_i(i=0,1,2)$ are given by
\begin{eqnarray}
m_0=-\frac{12k_1}{a},\quad m_1=m_2=6k_1\left(1-\frac{R_1^2}{a^2}\right).
\end{eqnarray}
Hence the total area of the two black rings is given by
\begin{eqnarray}
A_{\rm 2Ring}=-2ak_1\pi^2\sqrt{3l_1^2-48k_1^2\left(1-\frac{R_1^2}{a^2}\right)}.
\end{eqnarray}
The mass and the angular momenta depend on the parameters $k_1,l_1$ and $R_1$ as
\begin{eqnarray}
M \propto 8k_1^2+al_1,\quad J\propto -k_1\left[16k_1^2+3al_1+\frac{3}{2}(a^2-R_1^2)\right],
\end{eqnarray}
respectively. For fixed asymptotic charges $M$ and $J$, the parameters $k_1$ and $l_1$ are specified as functions of $R_1$. The partial derivative of $M$ and $J$ with respect to $R_1$ are computed as
\begin{eqnarray}
&&0=\frac{\partial M}{\partial R_1} \propto 16k_1\frac{\partial k_1}{\partial R_1}\biggl|_{M,J}+a\frac{\partial l_1}{\partial R_1}\biggl|_{M,J}, \\
&&0=\frac{\partial J}{\partial R_1}\propto-\left[48k_1^2+3al_1+\frac{3}{2}(a^2-R_1^2)\right]\frac{\partial k_1}{\partial R_1}\biggl|_{M,J}-3ak_1\frac{\partial l_1}{\partial R_1}\biggl|_{M,J}+3k_1R_1.
\end{eqnarray}
Solving these equations, we obtain the derivatives of $k_1$ and $l_1$ as follows 
\begin{eqnarray}
&&\frac{\partial k_1}{\partial R_1}\biggl|_{M,J}=\frac{-2k_1R_1}{R_1^2-a^2-2al_1},\\
&&\frac{\partial l_1}{\partial R_1}\biggl|_{M,J}=\frac{32k_1^2R_1}{a(R_1^2-a^2-2al_1)}.
\end{eqnarray}
Therefore, the partial derivative of the total area of two black rings with respect to $R_1$ is given by
\begin{eqnarray}
\frac{\partial A_{\rm 2Rings}}{\partial(-R_1)}\biggl|_{M,J}=\frac{4\sqrt{3}\pi^2k_1^2 R_1(24(a^2-R_1^2)k_1^2-a^2l_1^2)}{(a^2-R_1^2+2al_1)\sqrt{a^2k_1^2l_1^2-16(a^2-R_1^2)k_1^4}}.\label{eq:ratio}
\end{eqnarray}

For example, we consider the case where two black rings are put near the nuts at the north pole and the south pole on an $\rm S^2$-bolt, i.e., $-R_1=R_2\simeq a$. The right hand side of Eq.(\ref{eq:ratio}) is positive. Hence we should not expect the transition from black holes into black rings to occur. Next, we arrange that two black rings are located in the neighborhood of the equator on the $\rm S^2$-bolt, i.e., $-R_1=R_2\simeq 0$. In the case of $-4k_1<l_1<-2\sqrt{6}k_1$ it can take a negative value. Therefore we should not expect two black rings near the equator of the $\rm S^2$-bolt to spontaneously coalesce and change into a single black ring.


\appendix

\section{Eguchi-Hanson space}\label{sec:EH}

The metric of the Eguchi-Hanson space is given by
\begin{eqnarray}
&& ds^2_{\rm EH} = \left( 1- \frac{a^4}{\bar r^4}\right) ^{-1} d\bar r^2 
        + \frac{\bar r^2}{4} \left[ \left( 1-\frac{a^4}{\bar r^4} \right) 
          \left( d\bar\psi+\cos\bar\theta d\bar\phi \right)^2
        + d\bar\theta^2+\sin^2\bar\theta d\bar\phi^2 \right],
\label{metric_EH}
\end{eqnarray}
where $a$ are a constant, $0\leq \bar\theta\leq \pi,~ 
0\leq \bar\phi\leq 2\pi/$ 
and $0\leq \bar\psi\leq 2\pi$. 
The Eguchi-Hanson space has an S$^2$-bolt at $r=a$, 
where the Killing vector field $\partial/\partial \bar \psi$ vanishes. 

In order to clarify the relationship between the Gibbons-Hawking coordinate and the metric (\ref{metric_EH}), we introduce the coordinates as follows
\cite{Prasad},
\begin{eqnarray}
& & r=a\sqrt{\frac{\bar r^4}{a^4}-\sin^2\bar\theta},\quad 
\tan\theta=\sqrt{1-\frac{a^4}{\bar r^4}}\tan\bar\theta,\quad 
\phi=\bar\psi,\quad 
\psi=2\bar\phi. \notag \\
& &
( 0\leq \theta\leq \pi,~~ 
  0\leq \phi\leq 2\pi,~~
  0\leq \psi\leq 4\pi )
\end{eqnarray}
Then, the metric takes the form of 
\begin{eqnarray}
& & ds^2_{\rm EH} = V ^{-1} (r,\theta) 
               \left[ dr^2 + r^2 \left( d\theta^2 + \sin^2\theta d\phi^2 \right) \right]
               + V(r,\theta) \left( \frac{a}{8} d\psi+\varphi_\phi d\phi \right)^2,
\label{metric_GH}
\\
& & V^{-1} (r,\theta) = \frac{a/8}{|{\bm r}-{\bm r}_1|}
            +\frac{a/8}{|{\bm r}-{\bm r}_2|},\\ 
& &\varphi_{\phi}(r, \theta)
= \frac{a}{8} 
        \left(
  \frac{r \cos\theta - a}{\sqrt{r^2 + a^2 - 2 a r \cos\theta}}
+ \frac{r \cos\theta + a}{\sqrt{r^2 + a^2 + 2 a r \cos\theta}}
\right),
\end{eqnarray}
where ${\bm r}=(x,y,z)$ is the position vector on 
the three-dimensional Euclid space and ${\bm r}_1=(0,0,a)$, 
${\bm r}_2=(0,0,-a)$. 
The metric (\ref{metric_GH}) is the  
Gibbons-Hawking two-center form of the Eguchi-Hanson 
space\cite{Prasad,Eguchi,GH}. 
It is manifest in the coordinate that the space has 
two nut singularities at $\bm r= \bm r_j$ 
where the Killing vector field $\partial /\partial \psi$ vanishes.


\end{document}